\documentclass[aps,pra,twocolumn,showpacs,floatfix,groupedaddress]{revtex4-1}

\usepackage{eurosym}
\usepackage{graphicx}
\usepackage{subfigure}
\usepackage{color}
\usepackage{soul}
\usepackage{hyperref}
\usepackage{notoccite}
\usepackage{textcomp}
\usepackage{bigints}
\usepackage{amsmath}
\usepackage[normalem]{ulem}

\begin{document}

\title{Unidirectional Photonic Circuit with Phase-Change Fano Resonator}
\author{Roney Thomas, Eleana Makri, Tsampikos Kottos}
\address{Department of Physics, Wesleyan University, Middletown CT-06459, USA}
\author{Boris Shapiro}
\affiliation{Technion - Israel Institute of Technology, Technion City, Haifa 32000, Israel}
\author{Ilya Vitebskiy}
\address{Air Force Research Laboratory, Sensors Directorate, Wright Patterson AFB, OH 45433 USA}

\date{\today}

\begin{abstract}
We demonstrate that the integration of a phase-change material (PCM) in one of the two microresonators of a photonic 
metamolecule, coupled to a bus waveguide, can lead to unidirectional Fano resonances and to the emergence of a 
unidirectional transmission window. The phase change is caused by light-induced heating and is accompanied by an 
abrupt increase in the extinction coefficient of the PCM resonator. Due to the photonic circuit asymmetry, the critical 
value of the input light intensity triggering the phase change is strongly dependent on the input light direction. The 
latter determines the unidirectional nature of the emerging transmission window. This effect can be utilized in on-chip 
magnetic-free isolators and Q-switches.
\end{abstract}

\maketitle


\section{Introduction}
The realization of direction-dependent wave interference phenomena has become a compelling research topic with 
applications in electrodynamics, acoustics, matter waves, and quantum electronics \cite{ST91,JJWM08,D13,MEW13,
FSHA15,D05}. At the heart of this research effort is the technological demand for controllable asymmetric (e.g., 
unidirectional) wave transmission. On the fundamental side the challenge is to realize structures that violate the 
reciprocity principle in wave propagation. In the specific framework of photonics, such advances can lead to the 
realization of a new generation of optical isolators and circulators, unidirectional bio-sensors, switches, and 
modulators \cite{ST91,JJWM08}.
  
Traditional schemes that circumvent electromagnetic reciprocity via magneto-optical effects \cite{ZK97}, typically 
require strong bias magnetic field and/or a long optical path, none of which is compatible with on-chip integration. 
One possible solution is to use an active approach based on spatio-temporal modulation of the material parameters, 
such as the refractive index. The latter can be provided by electric, acoustic, mechanical, or optical dynamic biasing 
\cite{YF09,FYF12a,LYFL12,TFNFL14,FYF12b,ESSA14,KA14,SCA13}. At optical frequencies, though, it is hard to 
achieve spatio-temporal modulation of the refractive index, which would be strong enough for practical purposes. 
In addition, due to strong temporal modulation, the input and output signals might have different frequency composition 
for either direction of propagation. Alternative schemes involve spatially asymmetric nonlinear photonic structures 
\cite{RKGC10,LC11,BFBRCEK13,NBRMCK14,KA15,SSA18}. The nonlinear effects in transmission can be different 
for the forward and backward propagating wave, thus resulting in intensity-dependent propagation asymmetry. 
One problem with this approach is associated with the relative weakness of nonlinear optical interactions, such the 
Kerr effect.

In this paper we invoke a different approach for the realization of wave asymmetric phenomena and their underlying interference effects. Specifically, 
we consider light transport from a photonic circuit consisting of an asymmetric photonic metamolecule made by two micro-resonators side-coupled 
to a bus waveguide. One of the two resonators incorporates a phase-change material (PCM) which can be thermally driven to an insulator-to-metal 
(ITM) phase transition. For low irradiances of the incident wave, the PCM is in the insulating phase and the system shows a sharp Fano resonance 
\cite{JSK06,MFK10} irrespective of the direction of the incident wave. This Fano resonance is a consequence of wave interference effects between a 
high-Q resonant mode supported by the metamolecule and a scattering path existing within the continuum of states supported by the bus waveguide. 
When, however, the irradiance of the incident wave is above a critical value, one has to distinguish between two cases: (a) in the first case, the light 
is approaching the metamolecule from the side of the PCM resonator. In this case the PCM resonator undergoes a unidirectional abrupt increase of 
its extinction coefficient (associated with the presence of a nearby ITM phase transition) which spoils the high-Q of the resonant mode. In this case 
the Fano resonance is destroyed giving its place to a transparency window. If, on the other hand, the incident light enters the structure from the side
of the dielectric resonator, the Fano resonance remains intact. Specifically, a portion of the energy that the incident wave carries is lost (radiative losses) 
while it dwells in the dielectric resonator, thus reducing its efficiency to induce abrupt growth of the extinction coefficient of the PCM resonator which
remains in its dielectric phase. We find that this unidirectional Fano effect can lead to a transport asymmetry which is as high as $45dBs$ and has 
a frequency range which extends over several linewidths.

In this paper we invoke a different approach to the realization of strong asymmetry in wave propagation. Our approach does 
not rely on externally imposed modulation, magneto-optical nonreciprocity, or optical nonlinearity. Instead, we use the effect 
of self-induced heating on light transmission by a reciprocal but spatially asymmetric photonic circuit. In a free-space setting, 
similar mechanism of transmission asymmetry was considered in \cite{WHKSZZRGRGK18,ATKVK18}, where a plane wave 
was incident on an asymmetric multilayer with a phase-change component ($VO_2$). The PCM loaded layered structures in 
\cite{WHKSZZRGRGK18,ATKVK18} act as an asymmetric power limiter: it displays symmetric transmittance at low input light 
intensity, but the transmittance becomes highly asymmetric when the input power exceeds certain level. In this communication, 
we consider a qualitatively different setting, which involves a bus waveguide side-coupled to an asymmetric pair of microresonators 
(a photonic meta-
molecule). The first resonator incorporates a phase-change material (PCM) which can be thermally driven to an abrupt transition 
from the phase with lower absorption to the state with higher absorption. In our numerical example, the PCM is vanadium 
dioxide ($VO_2$), which undergoes insulator-to-metal (ITM) phase transition just above the room temperature \cite{Basov}. By contrast, 
the other resonator is characterized by a small but fixed extinction coefficient, which can be of the radiative nature. At low 
input light intensity, the PCM of the first resonator remains in the low-temperature, dielectric phase, and the system displays 
a sharp Fano resonance \cite{JSK06,MFK10}, thereby, blocking the wave propagation in either direction. The Fano resonance 
is a result 
of interaction between a high-Q resonant mode supported by the photonic meta-molecule and the continuum of states supported 
by the bus waveguide. The PCM resonator has some small initial absorption, which results in the light-induced heating. If the 
input light intensity exceeds certain threshold, it triggers the ITM phase transition in the PCM resonator. The important point is 
that this threshold is strongly dependent on the direction of light propagation. Specifically, if the input light approaches the 
meta-molecule from the side of the PCM resonator (hereinafter, the forward propagating light), the heating occurs faster, 
compared to the case when the light of the same intensity approaches the meta-molecule from the side of the second resonator 
(hereinafter, the backward propagating light). The abrupt increase in absorption, associated with the proximity of the ITM transition, 
spoils the high-Q of the resonant mode, thereby, suppressing 
the Fano resonance and giving its place to a transmission window. For the forward propagating light, the transition occurs at 
lower input light intensity than for the backward propagating light, implying the existence of a unidirectional transmission 
window above the ITM threshold. We find that this unidirectional Fano effect can lead to a transport asymmetry which is as 
high as $45dB$ and has a frequency range which extends over several resonance linewidths.

The structure of the paper is as follows: In the next section \ref{model} we analyze the transport properties of the phase-change 
photonic curcuit. Specifically, in subsection \ref{PCM} we present the photonic circuit and discuss the basic optical properties of 
the PCM resonator. In subsection \ref{comsol} we present the steady-state equations governing the transport properties of the 
system as well as the associated numerical scheme. In subsection \ref{ufano} we present our numerical results and analyze the 
interference mechanisms underlying the transport properties of the circuit for various input wave intensities. Specifically, we 
analyze the nature of the Fano interference and the way it disappears as the result of the ITM transition in the PCM resonator. 
Finally, in section \ref{theory} we present a simple theoretical model explaining the basic physics behind the unidirectional Fano 
resonances and the emergence of a unidirectional transmission window. Our conclusions are given at the last section \ref{conclusions}.

\section{Unidirectional Fano resonances in Phase-Change photonic circuits}\label{model}

In this section we present the transport characteristics of the phase-change photonic circuit. First (subsection \ref{PCM}), we describe the components 
of the photonic metamolecule that constitute our circuit and the optical properties of the PCM that is used. A typical optical PCM can be reversibly 
switched between two phases with different refractive index, optical absorption, or electrical conductance. The phase change can be caused by the 
input light itself (self-induced phase transitions) due to heating or some other physical mechanism~\cite{Cavalleri,Morin,Basov11}. Alternatively, it 
can be induced by external heating or cooling, by application of an electric or magnetic field, or by mechanical stress~\cite{Dressel, Maa, Locquet98}. 
Here we exclusively consider PCMs that undergo a self-induced phase transition due to the heating generated by the incident radiation.

In subsection \ref{comsol} we present the steady-state equations that describe the transport in the presence of a PCM while a more thorough analysis
of the numerical results is done in the following subsection \ref{ufano}. In this subsection we analyze the physical mechanism that leads to the presence 
of unidirectional Fano resonances. As a potential application, we show that such behavior can lead to asymmetric transport with asymmetry contrasts 
as high as $45$dBs. The unidirectional spectrally sharp Fano effect, and the associated strong field enhancement, can be utilized for a variety of other 
applications (i.e. apart from asymmetric transport) such as unidirectional low-threshold lasing, enhanced nonlinear response and (bio-)sensing. 

\subsection{Photonic circuit based on phase-change materials}\label{PCM}

\begin{figure}[h]
\includegraphics[width=1\columnwidth,keepaspectratio,clip]{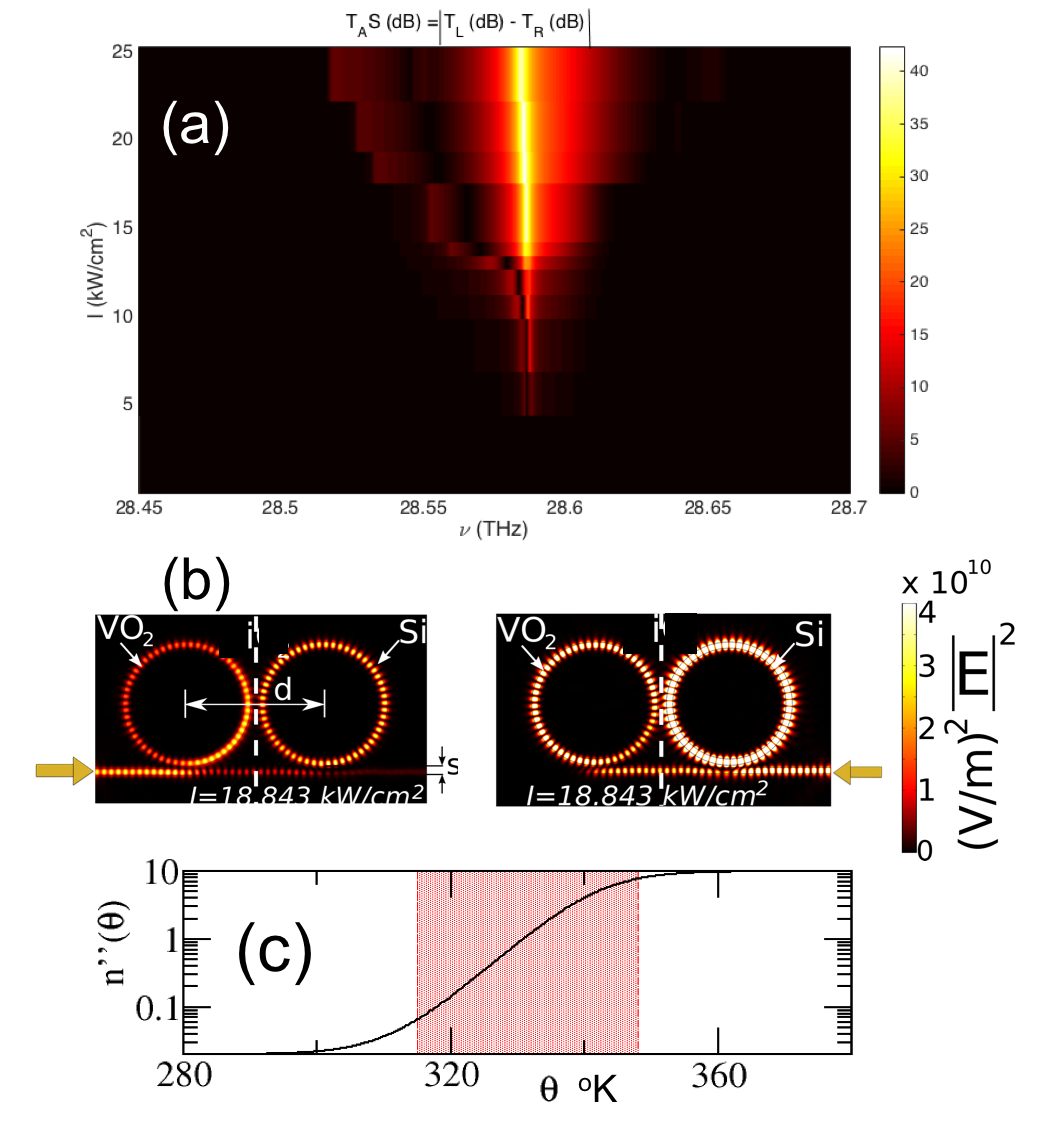}
\caption{(a) A density plot of left-right spectral transmittance asymmetry $T_L-T_R$ (in dBs) versus frequency $\nu$ (x-axis) of the incident CW, 
for various values of its irradiance ${\cal I}$ (y-axis). (b) A density plot of the scattering field intensity inside the circuit when it is illuminated with a 
CW signal from the left (right) direction. The frequency of the incident CW is at the Fano resonant frequency $\nu\approx 28.58 THz$ and its 
irradiance is $18.843kW/cm^2$. (c) The dependence Eq. (\ref{tempdepn}) of the imaginary part of the index of refraction $n^{\prime\prime}$ of the 
PCM from the temperature $\theta$. The red highlighted area indicates the temperature regime where the $Q$-factor of the PCM resonator is spoiled
leading to a destruction of the Fano mechanism.}
\label{fig1}
 \end{figure}

The proposed photonic metamolecule (see Fig. \ref{fig1}a) is designed to operate at the middle-infrared (MIR) regime of $10.5\mu m$. It consists 
of two ring resonators, side-coupled to a straight dielectric waveguide. The bus waveguide and the right ring resonator are made of lossless Si with 
refractive index $n_\text{Si} = 3.3$. The other (left) resonator is made of a PCM which has a temperature-dependent complex refractive index $n_\text{PCM} 
= n_{\text{PCM}}^{\prime} (\theta )+i n_\text{PCM}^{\prime\prime}(\theta )$. The thicknesses of the bus waveguide and the ring resonators are kept at 
1.7 $\mu$m. The center-to-center distance, $d$, between the two ring resonators, and between the rings and the straight waveguide, $s$, were kept 
fixed at $d= 35.8\mu \mathrm{m}$ and $s=2.295\mu \mathrm{m}$, respectively. The whole circuit lays on top of a ZnS substrate with refractive index 
$n_{\text{ZnS}}=2.2$.

The material used for the PCM ring is VO$_2$, which undergoes a phase transition from a monoclinic insulating phase to a rutile metallic phase around 
temperature $\theta_{c}=342$ K. When the temperature drops below $\theta_{c}$, the dielectric phase is restored and the material becomes optically 
transparent again. Furthermore, with the exception of a small hysteresis, only one of the two phases is stable at any given temperature: the dielectric phase 
is only stable below the phase transition temperature $\theta_{c}$, while the metallic phase is stable above $\theta_{c}$. The phase change transition 
is reflected in an abrupt variation of the imaginary part of the index of refraction $n_{\text{PCM}}^ {\prime \prime} (\theta)$ which can be as high as three 
orders of magnitude (depending on the deposition methods etc), see Fig. \ref{fig1}c. At the same time the real part of the index of refraction, 
$n_{\text{PCM}}^{\prime} (\theta) $ undergoes a relatively smooth variation. We have modeled these temperature driven variations in real and imaginary 
parts of the refractive index of $\text{VO}_{2}$ at the operational wavelength of $10.5\mu m$ by a direct fit of available experimental data found in the 
literature \cite{Kats13}. The best fit is represented by the following expressions
\begin{subequations} 
\label{tempdepn}
\begin{align}
n_{\text{PCM}}^{\prime}(\theta)=n_{0}^{\prime}+\frac{\Delta n^{\prime}}{\exp[-(\theta-\theta_{c})/\Delta\theta]+1}, \\
n_{\text{PCM}}^{\prime\prime}(\theta)=n_{0}^{\prime\prime}+\frac{\Delta n^{\prime\prime}}{\exp[-(\theta-\theta_{c})/\Delta\theta]+1}, 
\end{align}
\end{subequations}
where $n^{\prime}_0=3.3$ and $n^{\prime\prime}_0=0.001$,  $\Delta \theta=5 \text{K}$ denotes the smoothing parameter over which the phase transition 
takes place, and $\Delta n^{\prime} = 3.323$, $\Delta n^{\prime\prime} = 8.8$ indicate the ``heights'' of the jumps of the optical parameters during the 
phase transition. Below we will be considering the scenario for which the phase change from the dielectric to metallic state is caused by 
incident light-induced heating.

Finally, each of the ring resonators are designed to support a resonant mode at the mid-infrared wavelength of $10.5\mu \mathrm{m}$, corresponding 
to their optical mode number $m=27$. Using eigenmode analysis via Comsol multiphysics software, the quality factors of the resonant optical modes 
supported by the lossless silicon ring and the lossy PCM ring (in the insulating phase) are evaluated as $Q_\text{Si}=2.026\times 10^{4}$ and 
$Q_\text{PCM}=1.855\times 10^3$ respectively.

\subsection{Electromagnetic Transport Equations in the presence of Phase-Change Materials}
\label{comsol}

The electromagnetic wave propagation in the presence of a PMC component is a challenging computational problem \cite{TVK17,TVK18}. It involves 
a simultaneous solution of Maxwell's equations together with a heat-transfer equation which controls the temperature variation of 
the optical parameters of the PCM. In the steady state regime the problem collapses to the following set of coupled differential 
equations:
\begin{subequations}
\begin{align}
\nabla\times\vec{E}=i\mu_0\mu_r\omega\vec{H},\quad \nabla\times\vec{H}=-i\epsilon_0n^2(\vec{r},\theta)\omega\vec{E},  \\
-\nabla.(k(\vec{r})\nabla\theta(\vec{r}))=Q(\vec{r},\ \theta),  \label{heattransfer} \\
Q(\vec{r},\ \theta)=\epsilon_0\omega n^{\prime}(\vec{r},\ \theta)n^{\prime\prime}(\vec{r},\ \theta) \left|\vec{E}\right|^2
\label{heatlosses}. 
\end{align}
\end{subequations}
where $\vec{E}$ and $\vec{H}$ denote the electric and magnetic field vectors, $\mu_{0}(\epsilon_0)$ is the permeability (permittivity) of free 
space, $\mu_r=1$, and $n^{\prime}(\vec{r},\ \theta)$, $n^{\prime\prime}(\vec{r},\ \theta)$ are position and temperature-dependent functions 
that describe the real and imaginary part of the index of refraction inside the circuit. Furthermore, the parameter $k(\vec{r})$ in Eq. (\ref{heattransfer}), 
describes the thermal conductivity which, at the $\text{VO}_{2}$ ring, takes the value $k_{\text{VO}_{2}} =4\frac{W}{mK}$. We also 
assume that the ambient temperature at the edges of the surrounding $\text{ZnS}$ substrate is constant $\theta_0=293.15 K$. Finally, 
$Q$ (see Eq. (\ref{heatlosses})) is the heat production (per unit volume) which is generated at the lossy $\text{VO}_{2}$ ring resonator 
when the energy of the incident beam is dissipated there. The generated heat, in turn, leads to an increase in the temperature of the 
$\text{VO}_2$ ring resonator, which modifies its optical parameters $n^{\prime}(\theta)$ and $n^{\prime\prime}(\theta)$ as shown in 
Eqs. (\ref{tempdepn}). 

The steady-state values of transmittance $T_L/T_R$, phase transmission $\phi_L/\phi_R$ and temperature $\theta_L/\theta_R$, 
for left (L) and right (R) incident CWs (at various frequencies $\nu$) have been evaluated by solving numerically Eqs. (\ref{heattransfer}),
\ref{heatlosses}), using a Frequency-Stationary modulo of COMSOL MULTIPHYSICS. In order to maintain a high accuracy of our 
simulation results, we have used mesh elements of sizes 0.32 and 1.312 microns within the bus and ring resonator waveguides, and ZnS 
substrate. Throughout our simulations we have used a tolerance factor of 0.1\% as a convergence criteria. The accuracy of our calculation 
has been further checked by doubling the mesh density. Additionally port or reflectionless 
boundary conditions were used for launching and exiting electromagnetic waves within the model. 

In all our simulations below, we have not considered dispersion phenomena for $n^{\prime}(\vec{r})_{\rm PCM},n^{\prime\prime}
(\vec{r})_{\rm PCM}$. This approximation is justified by the fact that the variations of the index of refraction of the PCM due to 
temperature changes are much greater compared to any variations due to dispersion effects.

\subsection{Unidirectional Fano Interference and Asymmetric Transport}
\label{ufano}

The outcome of the simulations, have been analyzed and categorized according to the existence (or not) of sharp Fano resonance dips at
the transmission spectrum. We have found two distinct behaviors associated with the value of the irradiance of the incident CW: (a)  CW 
irradiances that support bi-directional (i.e. for both left and right incident waves) Fano resonance effects; (b) CW irradiances that support 
unidirectional Fano effects, i.e. the Fano resonance is suppressed when the incident CW enters the photonic circuit from a specific direction. 
As a result, the asymmetry contrast between left and right transmittances acquires high values in this regime.

\begin{figure}[t]
\includegraphics[width=1\columnwidth,keepaspectratio,clip]{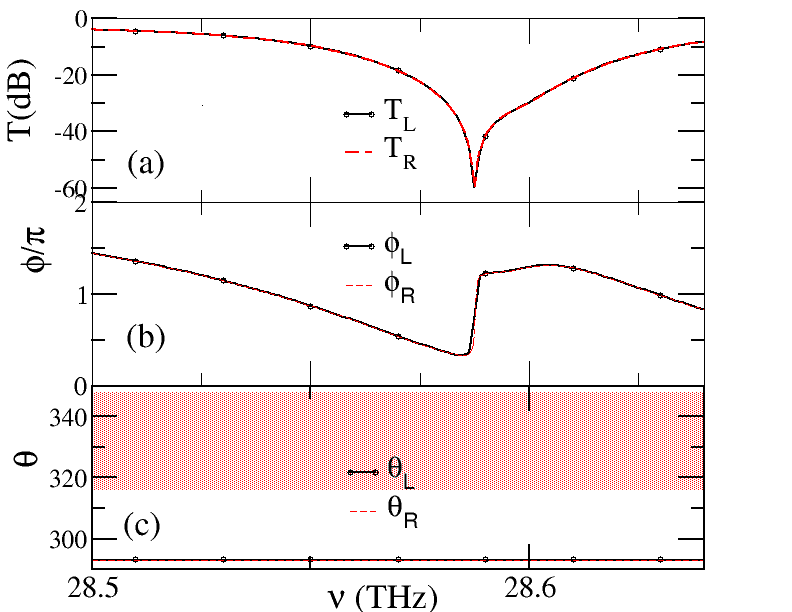}
\caption{(a) Transmittance $T_{L/R}$; (b) transmission phase $\phi_{L/R}$, and (c) temperature $\theta_{L/R}$ versus the frequency $\nu$ 
of an incident CW with irradiance $\mathcal{I}\approx 4.4 \text{kW/cm}^2$. Left (solid black lines)/right (dashed red lines) incident 
CW is indicated with the subindex L/R. The red highlighted area in subfigure (c) indicates the temperature domain where the variation of 
$n^{\prime\prime}_{\rm PCM}$ is approx. one order higher than its value in the dielectric phase. When the PCM temperature is in this range,
the Fano resonance is typically destroyed.
}
\label{fig2}
 \end{figure}

An overview of the left-right transmittance asymmetry, as we increase the irradiance ${\cal I}$ of the incident wave, is shown in Fig. \ref{fig1}a. 
We find that the left-right transmittance asymmetry increases, both in magnitude and in band-width, as ${\cal I}$ increases. In Fig. \ref{fig1}b 
we report (in a density plot) the scattering field inside the photonic circuit for a left and a right incident wave for a CW with moderate incident
irradiance ${\cal I}=18.843kW/cm^2$. We point out that higher values of the irradiance are physically irrelevant, since they correspond to 
electric field amplitudes of the incident wave that are above the electrical breakdown limit of the metamolecule. In our analysis below we 
ignore this regime and focus our analysis solely to low and moderate values of irradiances of the incident wave.

\begin{figure}[t]
\includegraphics[width=1\columnwidth,keepaspectratio,clip]{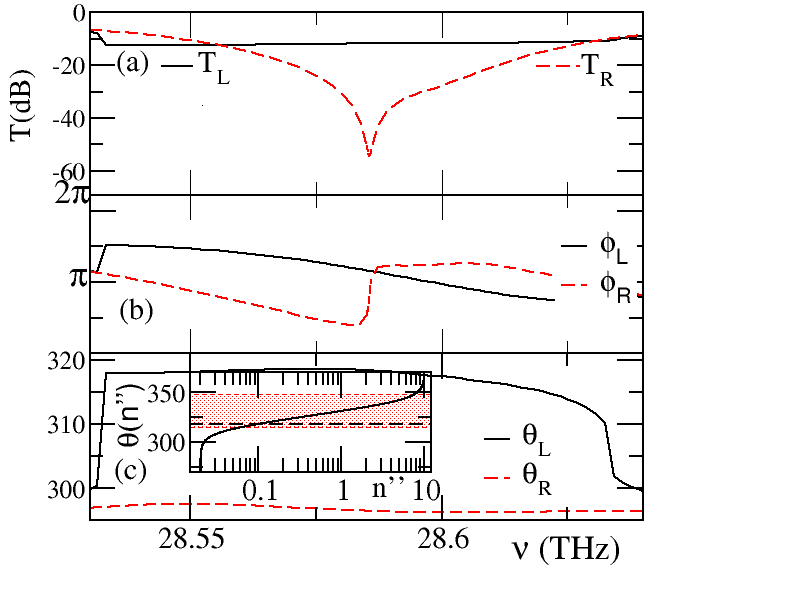}
\caption{(a)  Left $T_L$ (solid black lines) and right $T_R$ (dashed red lines) transmittances for a CW entering a photonic circuit 
(see inset) consisting of two coupled resonators with the left resonator made by a PCM material (VO$_2$). (b) Transmission phase 
for a left $\phi_L$ and right $\phi_R$ insident wave versus the incident frequency $\nu$; and (c) Associated temperature $\theta_L/
\theta_R$ for a left/right incident CW wave versus the incident frequency $\nu$. In the inset we show the variation of the imaginary 
part of the refractive index $n^{\prime\prime}$ versus the temperature $\theta$ of the PCM. The red highlighted area indicates the 
regime for which Fano resonances have been destroyed. The bold black dashed line indicates the value of $\theta\approx 318^0K$ 
associated with $\theta_L$ (see main panel of Fig. \ref{fig3}c). In all cases the irradiance of the incident CW is $\mathcal{I}\approx 
18.843 \text{kW/cm}^2$.
}
\label{fig3}
 \end{figure}

In Figs. \ref{fig2},\ref{fig3} we show typical results for CW with low, and intermediate irradiances respectively. Let us first consider 
the low irradiance case. From the numerical data shown at Fig. \ref{fig2} we see that in this case all transport parameters for left and right 
incident waves fall one on top of the other. Specifically, from the data shown in Fig. \ref{fig2}a,b we deduce that our metamolecule can 
support the formation of a sharp Fano resonance dip for both left and right incident waves at $\sim-60\mathrm{d}\mathrm{B}$ at $\nu 
=28.5872$. A calculation of the steady state temperature at the PCM resonator indicates that it remains much below $\theta_c$ for both 
left and right incident waves, see Fig. \ref{fig2}c. The existence of the Fano resonance in similar type of configurations has been well 
documented in the literature and therefore is not 
a surprise that a similar phenomenon occurs in our system as well. Its origin is traced back to the presence of fragile interference effects 
between two (or more) scattering paths being present in our photonic circuit: one of them involves a resonant process during which the 
incident wave is exciting a high-Q resonant (quasi-bound) state of the metamolecule while the other scattering path involves a state within 
the continuum which is supported by the bus waveguide. Traces of these interference effects can be found in the transmittance $T$ and 
the transmission phase $\phi$ when they are plotted versus the incident frequency $\nu$, see Fig. \ref{fig2}a,b. Specifically, in the proximity 
of the bound state frequency these quantities undergo abrupt variations, with the phase changing by $\pi$ within a spectral interval proportional 
to the resonance linewidth. At the same time the left and right transmittance profiles are equal $T_L=T_R$ and they acquire a non-Lorentzian 
asymmetric shape due to the destructive interference between the two paths.

The situation is completely different for intermediate values of the irradiance of the incident wave, see Figs. \ref{fig3}a,b,c. This regime is 
associated with irradiances which can induce temperature increase of the PCM material in a range around $\theta_c$. In such cases, the 
directionality of the incident wave might affect dramatically the fragile Fano interference. Specifically, due to the asymmetric design of the 
photonic circuit, a left or right propagating wave might encounter different losses before engaging with the PCM resonator. As a result they 
will carry different amount of energy and thus their efficiency to create a sizable increase of $n^{\prime\prime}_{\rm PCM}$ (or even to induce 
an insulator to metal transition) at the PCM resonator will be different.

Let us discuss in more detail the transport scenario in each of the cases associated with right and left incident CWs. Consider, for example, 
the situation where the CW enters the circuit from the side of the Si resonator (right incident direction). In this case, the light originally is 
trapped inside the Si resonator (see Fig. \ref{fig1}b) for a long time. Before it will be coupled back to the bus waveguide a portion of its carried 
energy is already lost due to the radiative losses from the Si resonator. As a result, the remaining energy is not enough to raise the temperature 
(via absorption) of the PCM resonator, and induce an increase of $n^{\prime\prime}_{\rm PCM}$, see Fig. \ref{fig3}c. Thus the latter remains 
in the dielectric phase, and the involved Fano interference mechanisms lead 
to a transport behavior which is similar to the one found in the case of incident waves with low irradiances. From Fig. \ref{fig3}a, we see that 
$T_R$ excibits an asymmetric resonance lineshape at approximately $\nu\approx 28.58$THz. At this frequency the transmission phase 
$\phi_R$ undergoes an abrupt change by $\pi$; a signature of Fano interference. Finally, the temperature $\theta_R$ at the 
PCM resonator remains well beyond the critical temperature $\theta_c$, see Fig. \ref{fig3}c. As a consequence the index of refraction 
(specifically $n^{\prime\prime}$) remains essentially constant and equal to the $VO_2$-value at the dielectric phase (see inset of this figure).

On the other hand, when the 
incident CW enters the photonic circuit from the side of the $VO_2$ resonator, its efficiency to heat up the PCM to temperatures that enforce 
sizable increases of the refraction index $n^{\prime\prime}_{\rm PCM}$ (say by an order from the dielectric value) remains intact. Such 
variations of  $n^{\prime\prime}$ spoil the high-Q resonant mode of the metamolecule. Consequently, the scattering path associated with the 
resonant mode is suppressed together with the Fano interference occurring between this path and scattering paths supported within the 
continuum of states from the bus waveguide. This leads to the emergence of a transparency window and an asymmetric transport between 
the left and right transmittance which can be as high as $45dBs$. The destruction of the Fano interference mechanism is also reflected in the
transmission phase $\phi_L$ which varies smoothly across the resonant frequency (Fig. \ref{fig3}b). In this frequency range, the temperature 
at the PCM resonator, 
has been raised to values $\theta_L\approx 318^0K$. Although this value is below the phase transition temperature $\theta_c$ it is, nevertheless, 
high enough to induce a variation in $n^{\prime\prime}_{\rm PCM}(\theta)$ of the PCM which is approx. an order of magnitude larger than 
the corresponding value of the insulating phase (see inset of Fig. \ref{fig3}c where $\theta_L\approx 318^0K$ is indicated with bold dashed black 
line inside the red highlighted area).

\section{Semi-analytical model based on coupled oscillators with a temperature-dependent damping coefficient}\label{theory}

\begin{figure}[t]
\includegraphics[width=\columnwidth,keepaspectratio,clip, trim= 0 4cm 0 4cm]{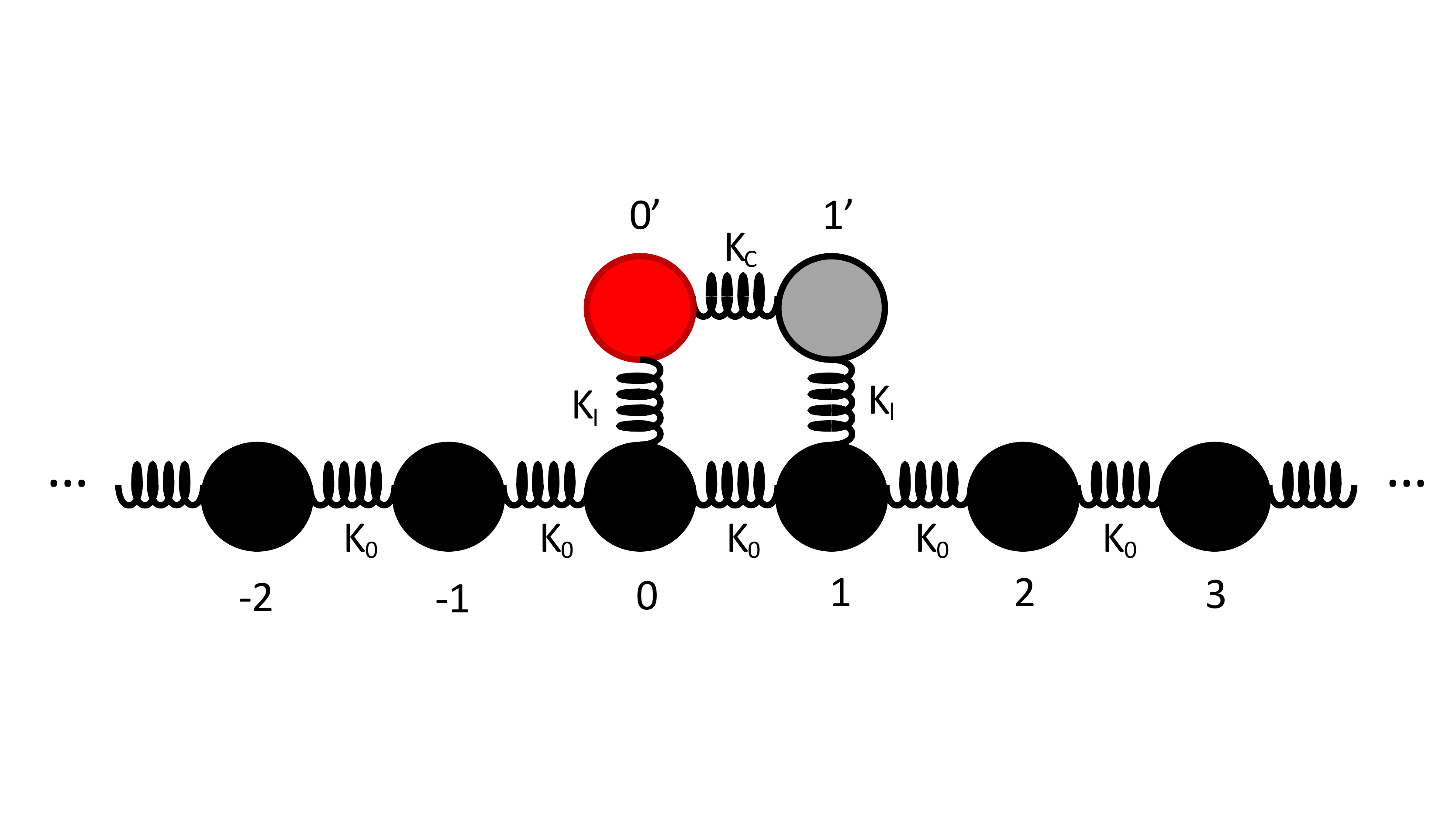}
\caption{Mechanical oscillator analogue of the photonic circuit of Fig. \ref{fig1}. Two coupled oscillators, one of which has temperature-dependent 
friction (red), are side coupled to an infinite array (lead)  of identical masses coupled with identical springs.}
\label{fig4}
 \end{figure}

In order to obtain a quantitative understanding of the asymmetric transport hosted by the photonic circuit of Fig. \ref{fig1}, we analyze a simple
model consisting of coupled oscillators. In particular, we model the bus waveguide as an infinite chain of identical masses $m$ coupled by 
springs with constant $K_0$ (see Fig. \ref{fig4}). The ring resonators are represented by two oscillators of equal mass, each of which 
is coupled with one mass in the infinite array with a spring of constant $K_l$. The two masses are also coupled to each other with spring 
constant $K_c$. The resonator on the left is assumed to have losses due to friction $\mu$. Later, we are going to incorporate a temperature 
dependence in analogy with the temperature dependent optical parameters of the VO$_2$ resonator in the photonic circuit of the previous section. 
The system is described by the following set of equations 
\begin{subequations}
\begin{align}
\ddot{X}_n &= - \omega_0^2 (2 X_n-X_{n-1} -X_{n+1})-\omega_l^2 (X_n-X_{0^\prime})\delta_{n,0}\nonumber\\
&-\omega_l^2 (X_n-X_{1^\prime})\delta_{n,1}, \label{inflead2} \\
 \ddot{X}_{0^\prime}&= - \omega_c^2 (X_{0^{\prime}}-X_{1^\prime}) - \omega_l^2 (X_{0^{\prime}}-X_{0})- \gamma \dot{X}_{0^\prime}, \label{lossyres2}\\
 \ddot{X}_{1^\prime} &=-\omega_c^2 (X_{1^{\prime}}-X_{0^\prime}) -\omega_l^2 (X_{1^{\prime}}-X_{1}), \label{sires2}
\end{align}
\end{subequations}
where $\omega_0=\sqrt{K_0/m}$, $\omega_l=\sqrt{K_l/m}$, $\omega_c=\sqrt{K_c/m}$, and $\gamma=\mu/m$ is the damping coefficient 
of the oscillator on site $0^\prime$. Above, Eq. (\ref{inflead2}) is the equation of motion describing the $n^{\text{th}}$ mass in the infinite 
array, Eq. (\ref{lossyres2}) is the equation of motion for the lossy resonator $0^\prime$, and Eq. (\ref{sires2}) is the equation of motion for 
the mass $1^\prime$. The radiative losses in this model are controlled by the coupling constant $k_l$ (though one could also introduce 
a small constant friction term at oscillator $n=1^{\prime}$ to imitate additional "radiative" losses).

Substituting the form $X_n (t)=a_n e^{-i \omega t}$ in the equations above we get
\begin{subequations}
\label{eqnmot}
\begin{align}
-\omega^2 a_n &= - \omega_0^2 (2 a_n-a_{n-1} -a_{n+1})-\omega_l^2 (a_n-a_{0^\prime})\delta_{n,0}\nonumber\\
& -\omega_l^2 (a_n-a_{1^\prime})\delta_{n,1}, \label{inflead} \\
-\omega^2 a_{0^\prime}&= - \omega_c^2 (a_{0^{\prime}}-a_{1^\prime}) - \omega_l^2 (a_{0^{\prime}}-a_{0})+ 
i \gamma \omega a_{0^\prime}, \label{lossyres}\\
-\omega^2 a_{1^\prime} &=-\omega_c^2 (a_{1^{\prime}}-a_{0^\prime}) -\omega_l^2 (a_{1^{\prime}}-a_{1}). \label{sires}
\end{align}
\end{subequations}
From Eq. (\ref{inflead}) we can easily find out that a propagating wave on the infinite chain has a dispersion relation 
$\omega=2 \omega_0 \sin (\frac{k}{2})$. Below we measure everything in units of $\omega_0=1$

We proceed with the analysis of the transport properties of the system of Fig. \ref{fig4}. To this end we consider an excitation through 
the infinite chain of coupled masses. For left incidence, the appropriate scattering boundary conditions are
\begin{subequations}
\begin{align}
a_n&= I_L e^{ikn}+R_L e^{-ikn},&n\leq0, \label{inc}\\
a_n&= T_L e^{ikn},&n>0. \label{ref}
\end{align}
\end{subequations}
Applying these boundary conditions to Eqs. (\ref{eqnmot}) using $n=0, n=1$ for Eq. (\ref{inflead}), we obtain the following system of 
equations
\begin{subequations}
\label{eqnmotions}
\begin{align}
\omega^2 (I_L+R_L) &= \omega_0^2 [ 2 (I_L+R_L)-I_L e^{-ik}-R_L e^{ik} - T_L e^{ik}]\nonumber\\
& +\omega_l^2 (I_L+R_L-a_{0^\prime}), \label{lead0} \\
\omega^2 T_L e^{ik} &= \omega_0^2 (2 T_L e^{ik} -I_L -R_L  -T_L e^{2ik})\nonumber\\
& +\omega_l^2 (T_L e^{ik}-a_{1^\prime}), \label{lead1} \\
\omega^2 a_{0^\prime}&= \omega_c^2 (a_{0^{\prime}}-a_{1^\prime}) + \omega_l^2 (a_{0^{\prime}}-(I_L+R_L ))\nonumber\\
& - i \gamma \omega a_{0^\prime}, \label{lossyresbc}\\
\omega^2 a_{1^\prime} &=\omega_c^2 (a_{1^{\prime}}-a_{0^\prime}) +\omega_l^2 (a_{1^{\prime}}-T_L e^{ik}). \label{siresbc}
\end{align}
\end{subequations}
which allows us (using Eqs. (\ref{lead0}, \ref{lead1})) to obtain the transmitted and reflected wave amplitudes in terms of the incident wave 
amplitude $I_L$, $\gamma$, and $a_{0'}, a_{1'}$:
\begin{subequations}
\label{TRL}
\begin{align}
T_L &=\frac{a_{0'} \omega_l^2 \omega_0^2+a_{1'}\omega_l^2(e^{-ik}\omega_0^2+\omega_l^2)-
2i I_L \omega_0^4 \sin k}{2 \omega_l^2 \omega_0^2 +e^{ik} \omega_l^4-2i\omega_0^4 \sin k},\\
R_L &=\frac{a_{0'} \omega_l^2 (\omega_0^2 +e^{ik}\omega_l^2) +a_{1'}\omega_l^2 \omega_0^2 e^{ik} -
I_L e^{ik} \omega_l^2(\omega_l^2+ 2 \omega_0^2 \cos k)}{2 \omega_l^2 \omega_0^2+e^{ik} \omega_l^4-2i\omega_0^4 \sin k}.
\end{align}
\end{subequations}
The associated transmittance from the left is $\mathcal{T}_L\equiv|T_L/I_L|^2$ and can be evaluated explicitly using 
Eqs. (\ref{TRL}, \ref{lossyresbc}, \ref{siresbc}). 

Using the same steps as above, we can calculate the transport for the case of a right incident wave. In this case the associated 
boundary conditions are
\begin{subequations}
\label{rscattering}
\begin{align}
a_n&= I_R e^{-ikn}+R_R e^{ikn},&n\geq 1, \\
a_n&= T_R e^{-ikn},&n<1.
\end{align}
\end{subequations}
which leads to the following expressions for the transmission and reflection amplitudes
\begin{subequations}
\label{TRR}
\begin{align}
T_R &=\frac{a_{0'} \omega_l^2 (\omega_0^2+ \omega_l^2 e^{ik})+a_{1'}\omega_l^2 \omega_0^2 e^{ik}-
2i I_R \omega_0^4 \sin k}{2 \omega_l^2 \omega_0^2 +e^{ik} \omega_l^4-2i\omega_0^4 \sin k},\\
R_R &=\frac{a_{0'} \omega_l^2 \omega_0^2 +a_{1'}\omega_l^2 (\omega_0^2 e^{-ik} +\omega_l^2)-
I_R  e^{-ik} \omega_l^2 (\omega_l^2+2 \omega_0^2\cos k)}{2 \omega_l^2 \omega_0^2+
e^{ik} \omega_l^4-2i\omega_0^4 \sin k}.
\end{align}
\end{subequations}
where $a_{0'},a_{1'}$ are obtained by solving the following equations
\begin{subequations}
\begin{align}
\omega^2 a_{0^\prime}&= \omega_c^2 (a_{0^{\prime}}-a_{1^\prime}) +\omega_l^2 (a_{0^{\prime}}-T_R )- 
i \gamma \omega a_{0^\prime}, \label{lossyresbcR}\\
\omega^2 a_{1^\prime} &=\omega_c^2 (a_{1^{\prime}}-a_{0^\prime}) +\omega_l^2 (a_{1^{\prime}}-I_R e^{-ik}-R_R e^{ik}). \label{siresbcR}
\end{align}
\end{subequations}
The associated transmittance from the right is $\mathcal{T}_R\equiv|T_R/I_R|^2$ and can be evaluated explicitly using 
Eqs. (\ref{TRR}, \ref{lossyresbcR}, \ref{siresbcR}).

Next we introduce the temperature dependence of the damping coefficient $\gamma$. The underlying assumption of the calculations 
below is the vality of an adiabatic approximation i.e. the heat release during one period of the oscillation is infinitesimally small. This allows 
us to assume that the relative change in the friction coefficient $\gamma(\theta)$ of the oscillator at site $0'$ (see Fig. \ref{fig4}) during 
one oscillation period is extremely small.

We use the following functional dependence of the friction coefficient $\gamma$ from the temperature 
\begin{equation}
\gamma (\theta) = \gamma_{min}+\frac{\gamma_{max}-\gamma_{min}}{\exp[-(\theta-\theta_c)/\Delta]+1},
\label{friction}
\end{equation}
where $\gamma_{max}$ and $\gamma_{min}$ are the maximum and minimum values of the damping coefficient respectively 
and $\Delta$ is a smoothing parameter. The functional form Eq. (\ref{friction}) is inspired 
by the temperature dependence of the imaginary part of the refractive index for the PCM material that was used in our optical 
simulations, see Eq. (\ref{tempdepn}a). For simplicity of the analysis, we do not consider here changes of the resonant frequency 
of the oscillator which would correspond to the change in real refractive index with temperature in the optical set-up.

  \begin{figure}[t]
\includegraphics[width=1\columnwidth,keepaspectratio,clip]{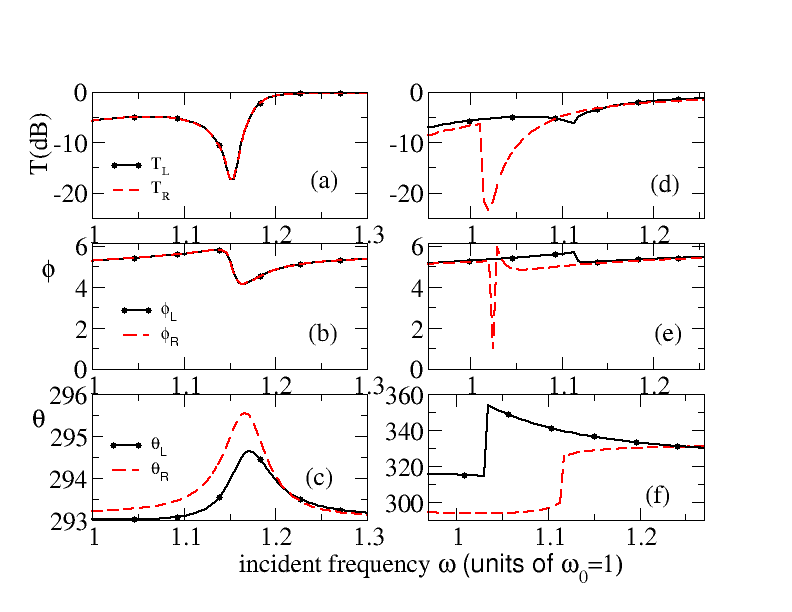}
\caption{Transport properties of the toy model of Eqs. (\ref{inflead2},\ref{lossyres2},\ref{sires2}).Left column: (a) Transmittance $T_{L/R}$; 
(b) transmission phase $\phi_{L/R}$, and (c) temperature $\theta_{L/R}$ spectra for a CW wave with field amplitude $0.5$. Left (solid black 
lines)/right (dashed red lines) incident signal is indicated with the subindex L/R. (d) $T_{L/R}$; (e) $\phi_{L/R}$ and (f) 
$\theta_{L/R}$ for an incident wave with field amplitude $3$. The transport behavior is in one-to-one relation with the results from the photonic 
circuit (see Figs. \ref{fig2} and \ref{fig3}).}
\label{fig5}
 \end{figure}

The change in temperature at the mass with friction involves two terms: (a) a heating term, which is the time averaged power dissipated 
due to friction that causes the heating and (b) a cooling term which describes the dissipation of heat. The heat rate equation is then
\begin{equation}
\label{rateeqn}
\frac{d\theta (t)}{dt}=-\kappa (\theta(t)-\theta_0)+\overline{P(t)},
\end{equation}
where $\kappa$ is the thermal conductance, $\theta_0$ is the ambient temperature. The dissipated averaged 
power over one period of an oscillation $\overline{P(t)}$ is
\begin{eqnarray}
\label{avpower}
\overline{P(t)}&= \gamma (\theta ) \langle \dot{X}_{0^{\prime}}^2 \rangle = \gamma (\theta ) \frac{1}{T} \int_0^T \Re \{X_{0^\prime} (t)\}^2 dt\nonumber\\
&= \frac{\gamma (\theta) \omega^2}{2} (\Re\{a_{0^{\prime}}\}^2+\Im\{a_{0^{\prime}}\}^2).
\end{eqnarray}
Substitution of the above expression in Eq. (\ref{rateeqn}) allows us to evaluate the steady-state temperature $\theta_{\infty}$ of the 
system of coupled oscillators, for a given incident wave amplitude $I_{L/R}$ and wavenumber $k$. Specifically, $\theta_{\infty}$ is
evaluated numerically as the root of the non-linear Eq. (\ref{rateeqn}) after imposing the steady-state condition $\frac{d\theta (t)}{dt}=0$
\cite{note1}.
The next step is to substitute $\theta_{\infty}$ in the expression Eq. (\ref{friction}) in order to evaluate the corresponding steady-state 
value $\gamma_{\infty}$. The latter is then used in Eqs. (\ref{TRL},\ref{TRR}) for the evaluation of the steady-state values of $\mathcal{T}_L, 
\mathcal{T}_R$.

Some representative cases of the steady-state values of left and right transmittances, transmission phases $\phi_{L/R}$ and temperatures 
$\theta_{L/R}$ of the oscillator $0^{\prime}$ vs. frequency for small $I_{L/R}=$ 0.5 (left column), and intermediate $I_{L/R}=$ 3 (middle column), 
values of the incident wave are shown in Fig. \ref{fig5}. The other parameters used are 
$\omega_0=1,\  \omega_c = \omega_l=0.7,\  \gamma_{min}=\kappa=0.02, \ \gamma_{max}=10, \ \theta_c=342,\  \theta_0=293,\  
\Delta = 5$. We focus our analysis on the antisymmetric resonance (the system shows another resonance for lower frequencies--associated
with a symmetric mode of the coupled dimmer oscillator). A direct comparison
with the transport results of the photonic circuit (see Figs. \ref{fig2},\ref{fig3}) indicates that our system shows the
same qualitative behavior. Specifically, there are two transport regimes associated with the incident field intensity: for low intensities
the system shows  bi-directional Fano resonances. In this regime the asymmetry in the left and right transmittances
is (if at all!) minimal. In contrast, when the intensity of the incident wave takes intermediate values the asymmetry becomes maximum. 
A simple inspection of Fig. \ref{fig5}c indicates that the origin of the asymmetric transport is the directional (i.e. from the left) heating of 
the oscillator $n=0^{\prime}$ and the consequent increase of its friction coefficient $\gamma$ which leads to suppression of the
Fano resonance, see Fig. \ref{fig5}a,b. One way to understand this asymmetric heating is by realizing that the transmitted energy towards
the oscillator $n=0^{\prime}$ involves only one scattering event at the junction $n=0$, while an incident wave from the right requires
at least two scattering events in order to reach the lossy oscillator.

Although the toy model that we have developed here is simple, it retains the basic physics principles that are responsible for the observed 
asymmetry in the transport properties of the photonics circuit of the previous section. It is analytically tractable and can potentially be explored 
further in order to better understand the transport near the phase-change transition point. In this case, the phase stability of the high-temperature 
regime and the hysteresis effects due to the heating and cooling cycles have to be carefully analyzed in order to obtain a better picture of 
the transport properties of our setup. The investigation along these lines will cast more light on the physics of photonic structures based on 
PCM and will be the subject of future work.

\section{Conclusions}\label{conclusions}

We have studied the transport properties of an asymmetric photonic metamolecule made of two micro-resonators side-coupled to a bus 
waveguide. One of them consists of a phase change material (PCM) and undergoes a phase transition from an insulating to a metallic 
phase due to self-induced heating by the incident radiation. For low incident irradiances, the metamolecule supports Fano resonances, 
associated with interference between a high-Q resonant mode of the metamolecule and a scattering state inside the continuum of states 
supported by the bus waveguide. In this case, Fano resonance is bi-directional and the transmittance asymmetry between left and right 
incident waves is minimal (if at all). For higher irradiances of the incident radiation, the Fano resonance is uni-directional. The phenomenon
is associated with the digration of the $Q$-factor of the resonant mode when the light enters the structure from the side of the PCM resonator.
We show that in this case the transmittance asymmetry is maximal and can reach values as high as $45dBs$. Our numerical
results for the optical photonic metamolecule are quantitatively captured by a simple mechanical model consisting of two coupled oscillators
where one of them has a temperature dependent friction coefficient.

{\it Acknowledgements --} (R.T., E.M., \& T.K.) acknowledge partial support from an ONR grant N00014-16-1-2803 and from DARPA 
NLM program via grant No. HR00111820042. (I.V.) was supported by an AFOSR FA9550-14RY14COR grant. The views and opinions 
expressed in this paper are those of the authors and do not reflect the official policy or position of the U.S. Government.


\begin{thebibliography}{99}
\bibitem{note1}For some parameter ranges, the heat rate equation has multiple roots thus exhibiting multistabilities. In these cases, we considered as the steady state solution the maximum temperature obtained that corresponded to a root and for which the second time derivative of the temperature is negative.
\end{thebibliography}
\end{document}